\documentclass[aps,showpacs,pra,amsmath,amssymb,twocolumn,superscriptaddress]{revtex4}
\usepackage{subfigure}
\usepackage{amsmath,amssymb,graphicx,hyperref}
\newcommand{\be}{\begin{equation}}
\newcommand{\ee}{\end{equation}}

\begin{document}

\title{Density, spin, and pairing instabilities in polarized ultracold Fermi gases}

\author{Inti Sodemann, D. A. Pesin, and A. H. MacDonald}
\affiliation{Department of Physics, University of Texas at Austin, Austin TX 78712 USA}

\date{\today}

\begin{abstract}
We study the influence of population imbalance on the pairing, spin, and density instabilities of a two component ideal Fermi gas after a sudden quench of interactions near a Feshbach resonance. Over a large region of parameters the pairing instability is dominated by finite momentum pairing, suggesting the possibility of observing FFLO-like states in the unstable initial dynamics. Long-wavelength density instabilities are found on the BCS side of the resonance, and are interpreted as a precursor of the phase separation expected at equilibrium. On the BEC side of the resonance, the pairing instability is present for scattering lengths that are larger than a critical value that is only weakly dependent on population imbalance and always smaller than the scattering length at which the Stoner-like spin instability occurs.
\end{abstract}


\maketitle

\section{Introduction}

Ultracold atom systems host a web of phenomena that have made it possible to
study quantum many body problems in new ways that are often remarkably revealing.
For example, studies of
fermionic atoms have enabled
a comprehensive investigation of the crossover between the BCS and BEC limits of the superfluid phase~\cite{crossover}.  Crossover physics studies have been further enriched
by exploring the fate of the superfluid phase as a function of the population imbalance between the species~\cite{expimbalance}
and there is now a fairly complete
understanding of the equilibrium phase diagram of this system~\cite{theoimbalance}.

One goal of ultracold fermion research is to shed light on the physics of interacting electrons.
Atomic systems can often be prepared with less uncontrolled disorder and more control over
the system parameters that determine interaction strengths, facilitating comparisons between theory and experiment.
However the fact that interactions between electrons are normally dominated by Coulomb
repulsion, whereas interactions between atoms are attractive at the densities of cold-atom systems,
can stand in the way of these quantum simulation efforts.

A particular challenge is the simulation of itinerant electronic magnetism, which is a combined consequence of strong repulsive interactions and Fermi statistics. The effective interaction strength between low-kinetic-energy, low-density
atoms is strongly repulsive only when their attraction supports a shallow bound state. This observation motivated an experiment~\cite{ketterle} which explored the possibility of driving ferromagnetic instabilities~\cite{Stoner} by using a Feshbach resonance to suddenly increase the effective repulsive interaction strength.
However, the analogy with electrons is incomplete, mainly because of the possibility to form weakly bound Feshbach molecules out of pairs of atoms. Indeed, recent experiments~\cite{ketterle2} and theoretical studies~\cite{Pekker,Zhang} indicate that molecule formation
is substantial even during the initial dynamics.

Motivated by these studies, in the present theoretical work  we study  the initial linearized dynamics of density, spin, and pairing fluctuations of a system of atoms containing two fermionic species with unequal populations that is prepared in an ideal Fermi gas state and placed near a broad Feshbach resonance.
We follow the approach of Pekker {\em et al.}~\cite{Pekker}, generalizing their work for the case of
balanced populations to the case of imbalanced ones.

We find that on the BCS side of the resonance the dominant instabilities are pairing and density phase separation instabilities.  We also observe that in the scattering-length/spin-polarization space FFLO finite-wavevector-pairing fluctuations are more prominent than FFLO states are in the equilibrium phase diagram. This suggests the possibility of observing FFLO-like features in the initial dynamics of an ultracold gas after a quench. Our results for the density instabilities on the BCS side are in qualitative agreement with previous studies of unstable collective modes in polarized systems~\cite{Lamacraft}. On the BEC side we find that the critical scattering length for the appearance of a long-wavelength spin-density instability remains larger than the critical scattering length to trigger the pairing instability. This finding indicates that a spin-density instability brought about by effective repulsions, which attempts to phase separate spin species, if observed, will necessarily be accompanied by substantial pair binding into Feshbach molecules even in the polarized gas.

Our paper is organized as follows.  In Sec.~\ref{pairingsec} we discuss the initial pairing instabilities
in a spin-polarized Fermi gas placed near a Feshbach resonance.  In Sec.~\ref{densitysec}
we turn to an analysis of longitudinal spin and density instabilities, which are coupled in spin-polarized systems. In Sec.~\ref{sectransverse} we present a qualitative analysis of the instabilities of the transverse spin densities.
Finally, we conclude in Sec.~\ref{summary} with a brief summary and discussion of our findings.

\section{Pairing instabilities}\label{pairingsec}

Near a broad Feshbach resonance the vacuum interaction between low-energy fermions that
are distinguished by a quantum number (referred to below as spin) is controlled only by their mutual scattering length.
As a consequence, any potential whose range is much smaller than the average interparticle separation
can be used as an effective potential to describe the physics near the resonance, provided it is tuned so as to
reproduce the physical scattering length.
The effective Hamiltonian of the system can be written in the form
\begin{widetext}
\be
H=\sum_{k,\sigma}\xi_{\sigma}(k) c^\dagger_{k\sigma}c_{k\sigma}+\frac{1}{2} \sum_{\sigma1,\sigma2}\int dr dr'v(r-r')\psi^\dagger_{\sigma1}(r)\psi^\dagger_{\sigma2}(r')\psi_{\sigma2}(r')\psi_{\sigma1}(r),
\ee
\end{widetext}
\noindent where $\xi_{\sigma}(k)=k^2/2m-\epsilon_{F\sigma}$, and $v(r-r')$ is an appropriate pseudopotential. By performing a ladder sum, the scattering amplitude in the Fermi sea (the {\em Cooperon}) can be expressed in terms of the scattering amplitude in vacuum and the occupation numbers of Fermi sea states~\cite{Pekker}:
\begin{widetext}
\be
\label{eq:cooperon}
C^{-1}(E,P)=\frac{m}{4 \pi }\Bigl(\frac{1}{a}+i\sqrt{m(E+\epsilon_{F\uparrow}+\epsilon_{F\downarrow})-\frac{P^2}{4}} \ \Bigr)+\int \frac{d^3q}{(2\pi)^3}\frac{n_{\uparrow\frac{1}{2}P-q}+n_{\downarrow\frac{1}{2}P+q}}{E-\xi_{\uparrow\frac{1}{2}P-q}-\xi_{\downarrow\frac{1}{2}P+q}},
\ee
\end{widetext}

\noindent
where $E=\omega_\uparrow+\omega_\downarrow$ is the total energy measured with respect to the sum of the chemical potentials of the scattering pair and $P=k_\uparrow+k_\downarrow$ is their total momentum. Particles of the same spin do not interact directly.

\begin{figure}
\includegraphics[scale=0.43]{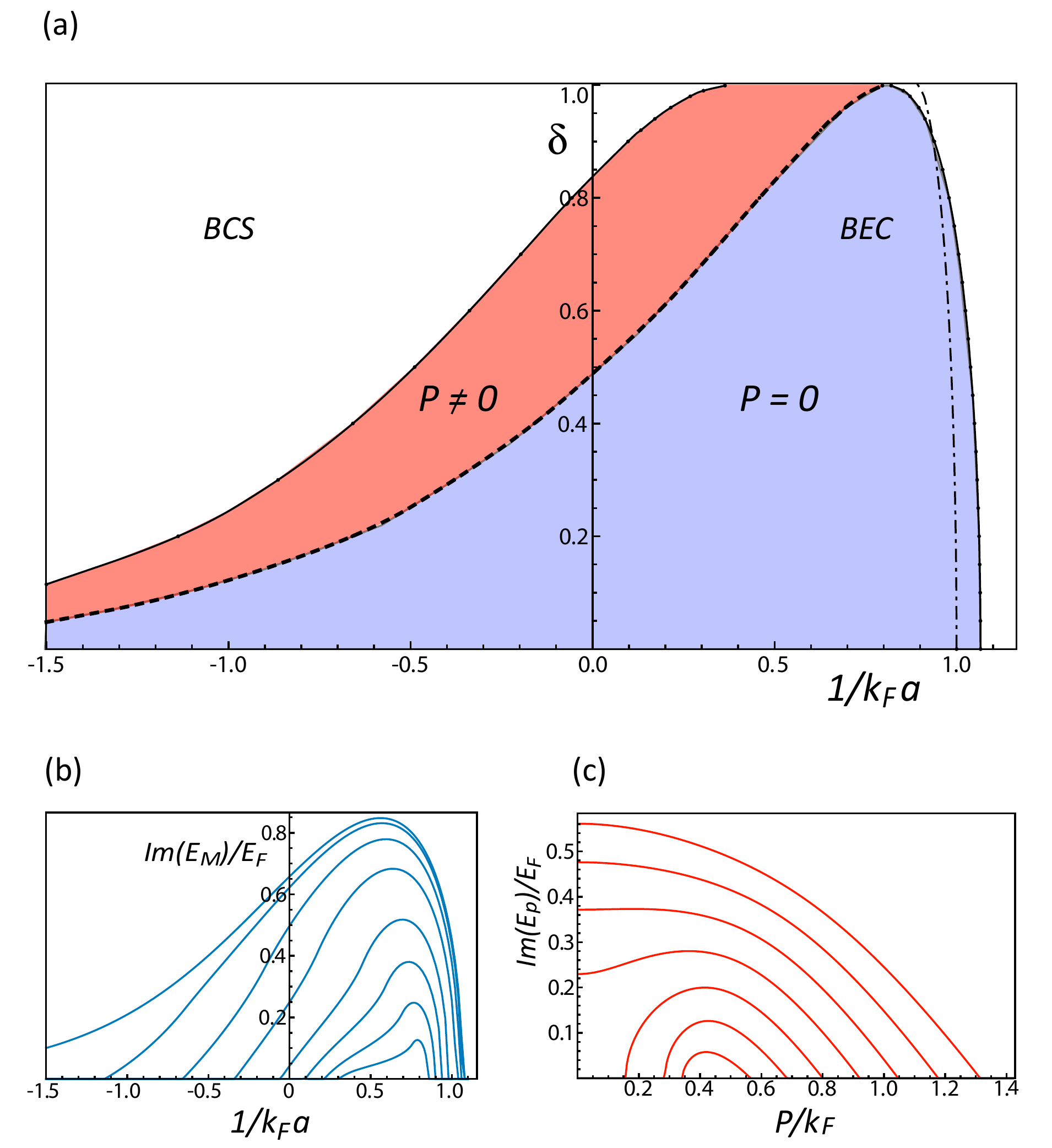}
\caption{(color online) (a) Pairing instability diagram. The red region in this diagram corresponds to those populations imbalances and scattering lengths at which the dominant pairing mode is at finite momentum, the blue region is that dominated by zero momentum pairing and in the white regions there are no unstable pairing modes. The dashed-dotted line is the approximate threshold for the Feshbach molecule formation limited by the energy of holes left in the Fermi sea discussed in the text, which nearly coincides with the onset of the pairing instability on the BEC side.
(b) Growth rate of the most unstable pairing mode $\rm{Im}(E_{M})$ as a function of the inverse scattering length across the resonance. The curves correspond to populations imbalances $\delta=\{0,0.2,0.4,0.6,0.8,0.9,0.96,0.99\}$.
Pairing instabilities occur over a narrower $k_Fa$ region at larger polarizations.
(c) Typical growth rates of the unstable pairing modes as a function of the pairing momentum, for $\delta=0.5$ and $1/k_F a=\{-0.4,-0.3,...,0.2\}$. $\rm{Im}(E_{P})$ is the imaginary part of the Cooperon pole energy for a given pairing momentum, these curves illustrate how the instability evolves from being dominated by pairing at finite momentum at negative scattering lengths to zero momentum at positive scattering lengths. We conjecture that quenches of the scattering length that approach the resonance from the BCS side might bring into realization states with substantial finite momentum pairing during the initial unstable dynamics.}
\label{pairing}
\end{figure}

Within time dependent mean-field theory it can be shown that the poles of the susceptibility associated with the pairing amplitude $\Delta_P=\sum_k \langle c^\dagger_{\frac{1}{2}P+k\uparrow} c^\dagger_{\frac{1}{2}P-k\downarrow}\rangle$ coincide with those of the Cooperon~\cite{Pekker, Pekker2}. The presence of poles in the upper half of the complex energy plane signals pairing modes that grow exponentially during the initial dynamics after a sudden quench of the interaction potential.
The equation $C^{-1}(E,P)=0$ thus determines the structure of unstable pairing modes, with the imaginary part of the pole
energy for a given momentum, ${\rm Im} (E_P)$, revealing the growth rate of the unstable mode in question.

The implications of this criterion for pairing instabilities in initial dynamics are summarized in
Fig.~\ref{pairing}. We have parameterized the interaction strength by the dimensionless product $k_{F}a$, where we define
$k_{F}$ for a polarized gas as the Fermi momentum of an unpolarized gas with the same total density,
{\em i.e.} $2 k_F^3=k_{F\uparrow}^3+k_{F\downarrow}^3$. In the latter expression $k_{F\uparrow/\downarrow}$ are the Fermi momenta of the two species of fermions.  We also define a population imbalance parameter
\be
\delta=\frac{n_{\uparrow}-n_{\downarrow}}{n_{\uparrow}+n_{\downarrow}}
=\frac{k_{F\uparrow}^3-k_{F\downarrow}^3}{k_{F\uparrow}^3+k_{F\downarrow}^3},
\ee

\noindent Fig.~\ref{pairing} illustrates how the fastest growing pairing modes depend on these two parameters.

For an unpolarized gas the pairing instability is present at all negative scattering lengths
(on the BCS side of the resonance), and grows most
rapidly for pairs with zero total momentum.
Interestingly, the instability survives over a finite region on the
BEC side of the Feshbach resonance
where the scattering length is positive.
The absence of an instability for small positive scattering lengths signals the metastability of the gas
with respect to its pair-fluctuation channels.
For a polarized gas we find that the instability can be dominated by finite momentum pairing.
This does not imply that the equilibrium state is a FFLO condensate of finite momentum pairs, but
does indicate that the initial dynamics of pairing modes is dominated by finite momentum pairing.

The large area of the region in the $k_{F}a$-$\delta$ phase diagram dominated by pairing at finite momentum
contrasts with the one associated with the FFLO phase in the equilibrium
phase diagram (The FFLO state is believed to occupy a very small region of the equilibrium phase diagram for a gas in three dimensions~\cite{theoimbalance}).  It falls roughly in the region of the equilibrium phase diagram
in which phase separation between superfluid and excess majority-spin
fermions is expected~\cite{theoimbalance}. Therefore, if a polarized system is prepared as a non-interacting gas on the BCS side far from the resonance, and the interactions are rapidly switched to the region where the
pairing instability exists, our results suggest the possibility of realizing states with substantial finite momentum-pairing during the initial dynamics. An important caveat to this picture is that the pairing instability is expected to compete with density instabilities associated with phase separation, as we will discuss in the next section.

When approached from the BEC side, the pairing instability onset is only
weakly dependent on spin imbalance.
The instability boundary can be interpreted as the threshold beyond which
the energy gained by molecule formation can be absorbed by adding two
holes to the Fermi sea~\cite{Pekker}. Approximating the binding energy as $\sim 1/m a^2$, the critical scattering length is found from
\be
\frac{k_{F\uparrow}^2}{2m}+\frac{k_{F\downarrow}^2}{2m}=\frac{1}{ma^2}.
\ee
\noindent This threshold is shown in Fig.~\ref{pairing} as a dashed-dotted line, and nearly coincides with the one of the onset of pairing instability on the BCS side. In particular, the maximum energy of two holes created in an unpolarized Fermi sea is $k^2_F/m$, whereas for the fully polarized Fermi sea it is $k^2_{F\uparrow}/2 m = 2^{-1/3} k^2_F/m$. Thus the ratio of the critical scattering length at which pair formation occurs in an unpolarized gas to the corresponding critical scattering length in a fully polarized (FP) gas is only slightly less than one,
\be
\frac{a_0}{a_{FP}}  \sim 2^{-1/6} \approx 0.89.
\ee
The actual ratio obtained directly from the Cooperon is $a_0/a_{FP}\approx0.75$, with $1/k_Fa_0\approx 1.07$ and $1/k_Fa_{FP} \approx 0.80$. 
When approached from the BCS side the pairing instability at full polarization
does not occur until the Feshbach resonance is crossed.
At full polarization, we can perform a similar estimate for the critical scattering length. The pair binding in this case occurs at finite momentum, between particles at the Fermi surface of the majority spin species and minority spin particles with zero momentum, so that the momentum of the pair is $P=k_{F\uparrow}$.
The energy of the pair is the sum of the binding energy and the center of mass kinetic energy, $E=-1/m a^2+P^2/4m$, while the holes left in both Fermi seas have vanishing energy since they lie near the Fermi surfaces. It follows that
\be
\frac{a_{FP}}{a^{FFLO}_{FP}}\sim\frac{1}{\sqrt{2}},
\ee
where we have referred to this critical scattering length as $a^{FFLO}_{FP}$ because the pairing occurs at finite momentum. We have $1/k_F a^{FFLO}_{FP} \approx 0.6$, which deviates considerably from critical value calculated directly from the Cooperon $1/k_F a^{FFLO}_{FP} \approx 0.4$. This discrepancy originates from its proximity to the unitary regime, where the molecular dispersion is greatly affected by the presence of the Fermi sea.

\section{Density and longitudinal spin-density instabilities}\label{densitysec}

We now turn to the instability analysis of density, $\rho_q=\sum_{k,\alpha}  c^\dagger_{k\alpha}c_{k+q\alpha}$, and spin-density along the polarization axis, $s_{zq}=\sum_{k,\alpha} c^\dagger_{k\alpha} \, \sigma^{\alpha\alpha}_z  \,c_{k+q\alpha}$. The analysis is again done by looking for the occurrence of poles with positive imaginary parts in the corresponding susceptibilities. The responses of these quantities are coupled in the polarized gas and share the same unstable collective modes. We defer the analysis of the unstable collective modes of the transverse spin-densities to Sec.~\ref{sectransverse}.

In time-dependent mean-field-theory, the spin and density dynamics after a sudden interaction-strength quench have a random phase approximation (RPA) form. In this recipe, the bare interaction potential enters the ``denominators'' of the RPA susceptibilities. Often the short-range interactions in a dilute gas are modeled with a contact pseudo-potential whose strength is proportional to the vacuum scattering length. This prescription, for which the RPA quenching picture is justified, would inevitably fail near the resonance. Thus a conflict arises in the description of the unstable dynamics of spin and density near the resonance. With this on mind we will examine density and spin instabilities by assuming the RPA form for the dynamics, but with the bare interaction replaced by the Cooperon. We believe that this approach, which has been previously employed to study spin instabilities in the unpolarized gas~\cite{Pekker,Pekker2}, is a reasonable compromise, but we do not know of a systematic justification for it.

The basic question we explore in this section is whether a free gas prepared far from the resonance can encounter a long-wavelength spin or density instability as the scattering length increases without
reaching the critical scattering length that determines the boundary
for the pairing instability.  If a spin-density instability that caused an increase
in the local spin-polarization occurred, it could inhibit pairing and allow the physics of
ferromagnetism to be studied in a cold atom context.  As discussed below,
our calculations indicate that when the resonance is approached from the BEC side the spin-density instabilities always occur within regions that are already unstable to pair formation. Even though we do not compare directly the growth rates of both instabilities, this finding is discouraging for the realization of an instability in which effective repulsive interactions would attempt to enhance locally the spin polarization of the gas.
\begin{figure}
\includegraphics[scale=0.85]{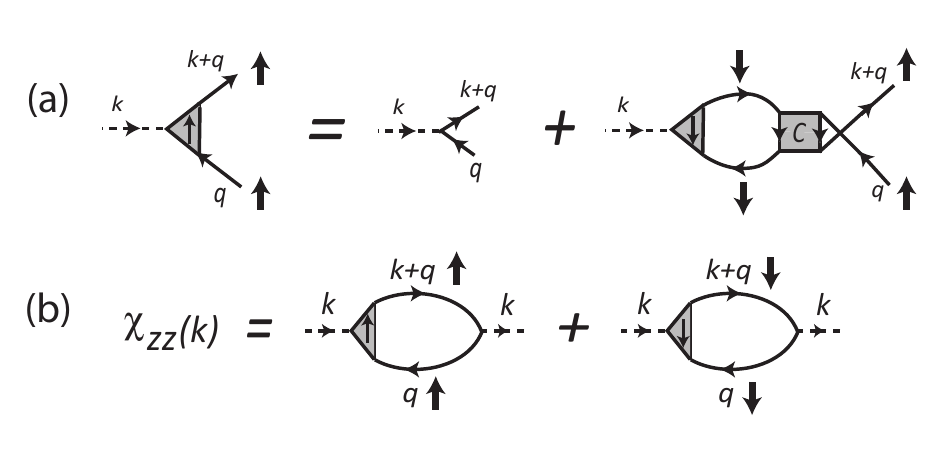}
\caption{a) Diagrammatic representation of the integral equation for the spin susceptibility vertex function for spin up.
b) Relation between the spin vertices and the longitudinal spin susceptibility.}
\label{diagrams}
\end{figure}

In a polarized gas the density and longitudinal spin responses are mutually coupled, but decoupled from the transverse spin response. The density and longitudinal spin susceptibilities can be expressed in terms of the spin-resolved density-density response functions~\cite{Vignale},

\be
i\chi_{\sigma\sigma'}=\langle T[\hat{n}_\sigma(r,t)\hat{n}_{\sigma'}(r',t')]\rangle.
\ee

\noindent In particular, $\chi_{zz}=\chi_{\uparrow\uparrow}-\chi_{\downarrow\uparrow}+\chi_{\downarrow\downarrow}-\chi_{\uparrow\downarrow}$ and $\chi_{\rho\rho}=\chi_{\uparrow\uparrow}+\chi_{\downarrow\uparrow}+\chi_{\downarrow\downarrow}+\chi_{\uparrow\downarrow}$. The collective modes can be conveniently studied by defining the density, $\Gamma^\rho_\sigma$, and spin, $\Gamma^s_\sigma$, vertex functions as follows:
\be
\chi_{\sigma\sigma}(k)\pm\chi_{\bar{\sigma}\sigma}(k)=-i\int_qG_{\sigma}(q)G_{\sigma}(k+q)\Gamma^{\rho,s}_\sigma(q,k),
\ee
\noindent where the $+ (-)$ sign corresponds to the density (spin) vertex, $q$ and $k$ are four component vectors that include frequency and momenta components, $G_{\sigma}(q)$ is the single-particle Green's function, and $\bar{\sigma}$ denotes the spin opposite to $\sigma$. The vertex functions satisfy the following self-consistent integral equations (see fig.~\ref{diagrams}):
\be
\label{eq:vertex}
\Gamma^{\rho,s}_{\sigma}(q,k)=1\mp i\int_{q'} G_{\bar{\sigma}}(q') G_{\bar{\sigma}}(q'+k) C(q+q'+k)\Gamma^{\rho,s}_{\bar{\sigma}}(q',k),
\ee
\noindent where $C$ is the Cooperon. The poles of the vertex functions determine the unstable collective modes. 

Integral equations of the above type are familiar~\cite{Abrikosov} from Fermi liquid theory.
In the limit of small wavelength $|k|<<k_F$ the dominant contribution to the integral over $q'$ in Eq.(\ref{eq:vertex})
is associated with low-energy Green's function poles. This property can be exploited~\cite{Abrikosov} by assuming that $\Gamma$ is slowly varying near
these poles.  This allows to substitute
\be\label{Gproduct}
G_\sigma(k)G_\sigma(k+q)\rightarrow \frac{4 \pi^3i}{k_{F\sigma}^2} \zeta_\sigma(q,\hat{k}) \delta(k_0-\epsilon_{F\sigma}) \delta(k-k_{F\sigma}),
\ee
where $\zeta_\sigma(q,\hat{k})$ is chosen so that its angular average
reproduces the long-wavelength, low-energy behavior of a
a free-particle (Lindhard) response function:

\be
\int \frac{d\Omega_{\hat{k}}}{4 \pi} \;  \zeta_\sigma(q,\hat{k}) = \chi^0_\sigma(q) = \int_k \frac{n_{k\sigma}-n_{k+q\sigma}}{\omega+\xi_k-\xi_{k+q}}
\ee

This approximation corresponds physically to
the idea that the initial dynamics of the ideal gas is dominated by
those states that are closest to it in energy, and that they in turn are formed from the gas state by making particle-hole
excitations that live near the Fermi surface.
The integral equation for the interaction vertices
can then be converted into a matrix equation involving the expansions of $\Gamma$,$\zeta$ and $C$ into spherical harmonics.
By truncating the resulting equations at the s-wave term
we obtain an expression for the vertices which resembles the RPA result,

\be\label{vertex2}
\Gamma^{\rho,s}_\sigma(k)=\frac{1\pm C_s(\omega)\chi^0_{\bar{\sigma}}(k)}{1-C^2_s(\omega)\chi^0_\uparrow(k) \chi^0_\downarrow(k)},
\ee

\noindent where $C_s(\omega)=\frac{1}{2}\int d(\hat{k}\cdot\hat{k'}) C(\omega,k_{F\uparrow}\hat{k}+k_{F\downarrow}\hat{k'}) $, is the s-wave average of the Cooperon. We expect the s-wave truncation to be reliable for the stability boundary estimate
at long wavelengths since in the limit of small $(\omega,q)$ we have

\be
\zeta_\sigma(q,\hat{k})\rightarrow \nu_\sigma\frac{\hat{q}\cdot\hat{k}}{\frac{\omega}{v_{F\sigma} q}-\hat{q}\cdot\hat{k}},
\ee

\noindent where $\nu_\sigma=mk_{F\sigma}/2\pi^2$ is the density of states at the Fermi surface for the spin $\sigma$. Assuming that the static susceptibility ($\omega/v_{F\sigma}q \rightarrow 0$) correctly predicts
the existence of long-wavelength instabilities, the only non-vanishing component in the spherical
harmonic expansion of $\zeta(q,\hat{k})$ in this limit is the s-wave term.
It can then be argued that the instability boundary is correctly estimated by the
s-wave truncation of Eq.~\eqref{vertex2}.

\begin{figure}
\includegraphics[scale=0.5]{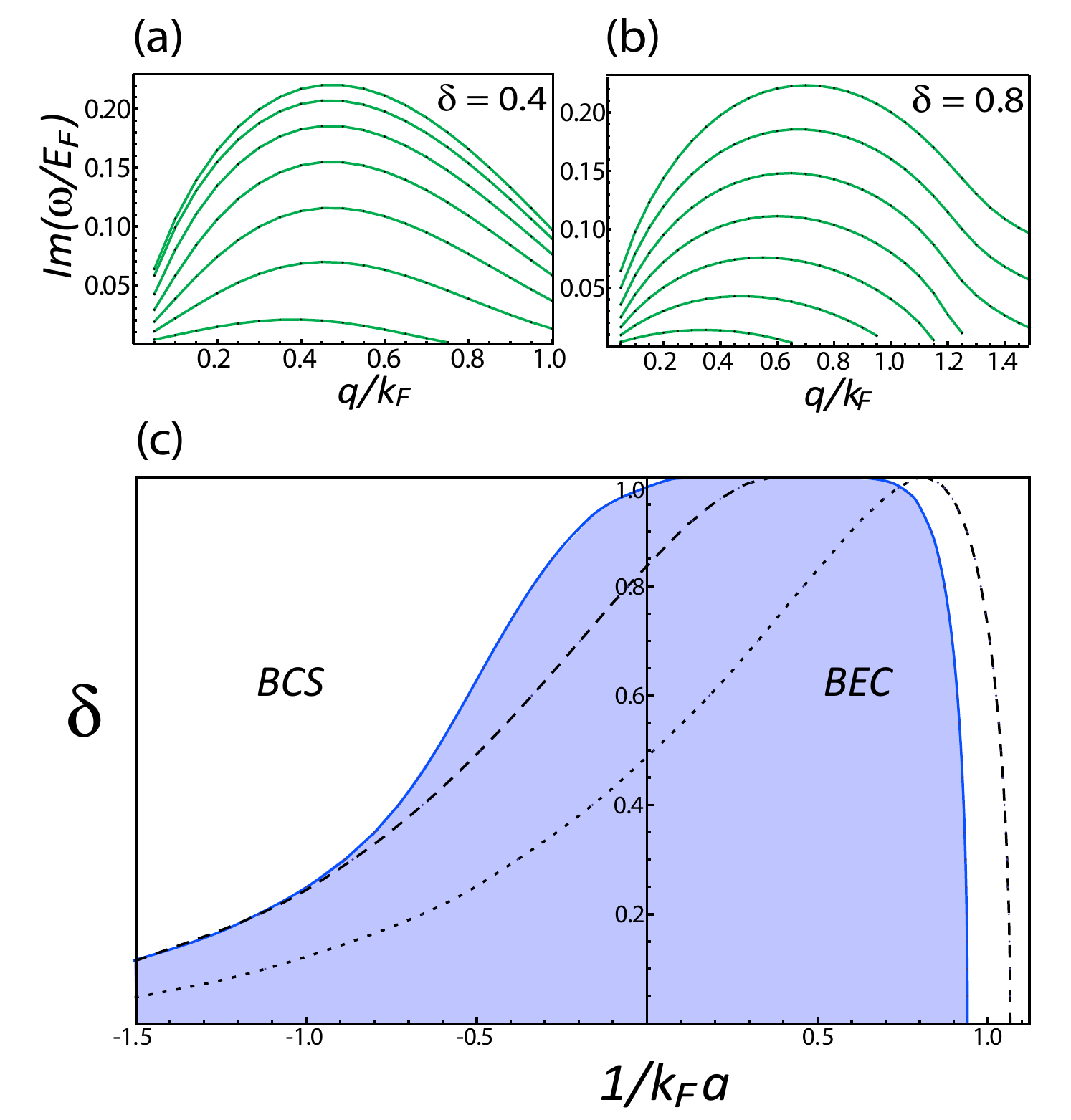}
\caption{(color online) (a) Growth rate of the density-longitudinal spin instabilities for $\delta=0.4$ and scattering lengths $1/k_F a = \{0.6,0.65,...,0.9\}$. (b) Growth rate of the density/spin instabilities for $\delta=0.8$ and scattering lengths $1/k_F a = \{-0.3,-0.25,...,0\}$. (c) Instability phase diagram.
The blue shaded region corresponds to the density-longitudinal spin instability region and for
comparison the dashed lines depict the boundary of the pairing instability region,
the dotted line separates the region dominated by finite momentum pairing from that dominated by zero momentum pairing (see Fig.~\ref{pairing}). When the resonance is approached from the BEC side the spin-density instability always appears at a scattering length where the gas is unstable to pairing as discussed in the text.}
\label{spindensity}
\end{figure}

Our results for the density and longitudinal spin-density instabilities in polarized gases are
summarized in Fig.~\ref{spindensity}.
We find that, when the resonance is approached from the BEC side, the pairing instability always occurs for smaller scattering lengths than the spin instability.
The spin-density instability occurs at smaller scattering lengths than those predicted
when the Cooperon is replaced by the contact pseudopotential, $C\rightarrow 4\pi a/m$, valid in the dilute limit, $k_F a\rightarrow 0$.  When combined with RPA, this replacement would imply a Stoner instability at $1/k_Fa=2/\pi$ for the unpolarized gas, whereas with the Cooperon we obtain $1/k_Fa\approx0.94$~\cite{Pekker}. The difference can be
understood as being due to the regularized scattering, which produces stronger repulsions when the resonance is approached from the BEC side. Comparable enhancements were found in Monte Carlo studies~\cite{Troyer,Trivedi}. Nevertheless, our results indicate that the population imbalance increases the critical
repulsion required to trigger spin instabilities, and this critical repulsion remains always larger than the corresponding one to produce pairing. We conclude that the preparation of a polarized initial state does not lead to
circumstances under which the Stoner instability of longitudinal spin modes is present, but the pairing instability is not.

Even though the instability boundaries for the spin-density instability can be calculated consistently within our approach, the growth rates of these instabilities are reliable only near the critical parameters at which the instability appears. Inside the unstable region, beyond a critical scattering length, we have found discontinuous behavior of the unstable modes as a function of momentum and scattering length. Since Eq.~\eqref{vertex2} is justified only at small frequencies and wave-vectors, it is difficult to judge the extent to which such behavior is reliable.  These discontinuities are also present in the density response of the unpolarized gas on the BCS side and are thus inherited by the longitudinal spin modes at finite polarizations because of their coupling to the density. We will therefore focus only on the instability boundaries in this work.

We first discuss the instability boundaries approached from the BCS side. As the resonance is approached at large polarizations, a long-wavelength density instability is always encountered at a lower scattering length than the corresponding
critical scattering length for igniting pairing instabilities.
At small polarizations ($\delta \lesssim 0.4$) this boundary becomes increasingly close to the one of the pairing instability. The  spin-density instability on the BCS side is brought about by effective attractive interactions. By examining the residue of the susceptibility pole, we have verified that this instability corresponds to an unstable mode in which the majority spin and minority spin densities grow with the same sign locally (that is, either they both decrease or both increase at any given point), with the amplitude of minority spin density larger than that of the majority spins.
In other words, this instability attempts to create an alternation of regions with small population imbalance
and regions with depleted minority spins. The relative amplitudes of the spin densities of the unstable mode near the instability boundary are,
\be
\delta n_\uparrow \propto \sqrt{\frac{\nu_\downarrow}{\nu_\downarrow+\nu_\uparrow}} \ , \  \delta n_\downarrow \propto \sqrt{\frac{\nu_\uparrow}{\nu_\downarrow+\nu_\uparrow}}.
\ee

We view this as the precursor of the creation of regions where the pairing instability is locally developed, alternated with regions where the minority spins are depleted, explaining why the pairing and density instabilities are so close together.
We expect that initial fluctuation bubbles with a growing density of paired particles should eventually merge and phase separate from the excess majority spin particles. This instability is therefore compatible with the equilibrium state, which has phase separated superfluid and normal gas regions, although our instability boundaries only indicate the appearance of linearly unstable modes and do not necessarily coincide with the boundaries between equilibrium phases.
\begin{figure}
\includegraphics[scale=1]{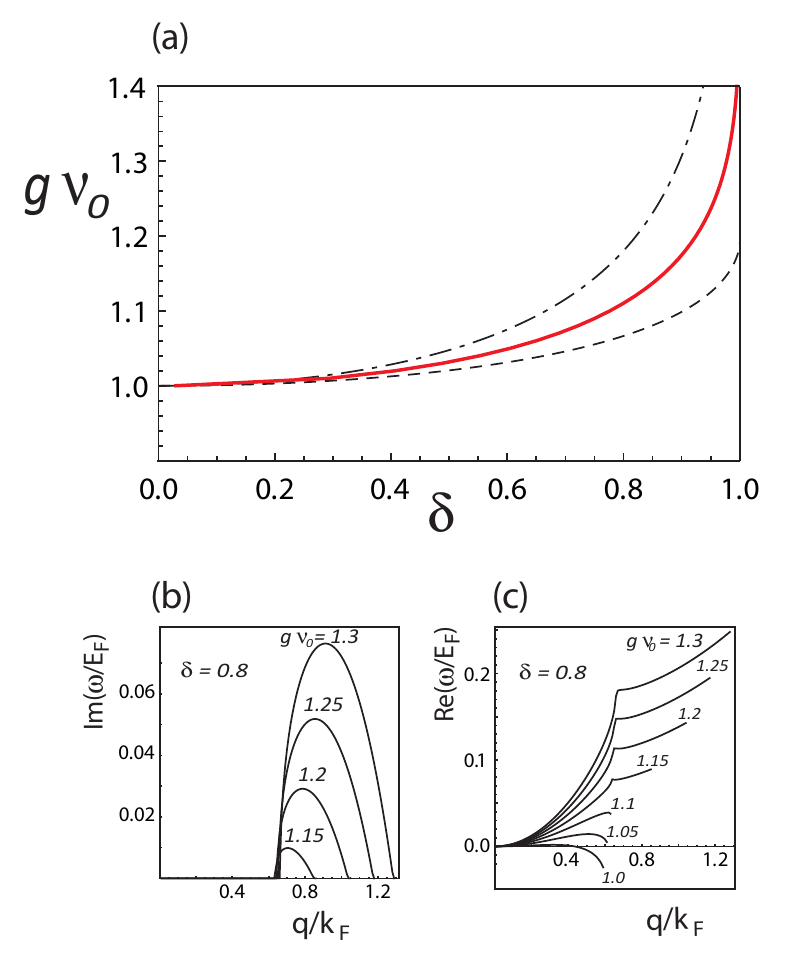}
\caption{\label{transverse}(color online) (a) critical repulsion strength for the onset of transverse spin instabilities (solid red line). The dashed line is the repulsion strength needed to sustain a spin imbalance in equilibrium ({\it i.e.} $\delta\mu=0$), which is finite at $\delta=1$. The dashed dotted line is the critical strength for the onset of longitudinal spin instabilities for the contact pseudopotential model. All lines converge to the Stoner criterion as $\delta\rightarrow0$. (b) Typical growth rates of the unstable modes (imaginary part of the poles) of the transverse spin instability. (c) Real part of the collective modes dispersion. The unstable modes evolve continuously from the conventional spin waves in equilibrium as explained in the text and illustrated in Fig.~\ref{spinwaves}.}
\end{figure}

We note that at smaller polarizations, $\delta \lesssim 0.4$, the boundaries of finite momentum pairing and density instabilities almost merge. In this regime we conjecture that sudden quenches might bring into realization states with substantial finite momentum pairing in the initial dynamics.

Several of our observations on the unstable dynamics on the BCS side are in qualitative agreement with a previous study of the unstable modes in the spinodal region of the polarized system~\cite{Lamacraft}. There, however, it was assumed that the dynamics of pairing was essentially instantaneous, so that the order parameter adjusted to its local equilibrium value after the quench, whereas the density and spin modes would respond on slower time scales. Although this is enforced at arbitrarily large wavelengths due to conservation laws, the dynamics of pairing and density-longitudinal spin modes near the resonance occurs on similar time scales $\sim \epsilon_F^{-1}$ at shorter wavelengths as illustrated in Figs.~\ref{pairing} and~\ref{spindensity}.

\section{Transverse spin instabilities}\label{sectransverse}

As discussed previously, near the resonance the momentum and frequency dependence of the effective two-particle interaction is important. As will become clear in this section, the instabilities of the transverse spin density occur at finite wavevectors, with the typical wavelength of the most unstable modes decreasing as the imbalance of populations increases. This situation complicates the analysis of the integral equations for the transverse spin vertices, since the Fermi liquid simplifications discussed in the previous are justified only at long wavelengths and small frequencies. Nevertheless, important insights into the nature of the transverse spin instability can be gained with the use of a frequency-momentum independent pseudopotential $V(r)= g \delta(r)$, and we will present them in this section. We leave open for future studies a more systematic treatment of this instability.

The response of the $x,y$ spin densities can be obtained in terms of the susceptibility $\chi_{\downarrow\uparrow}(q,\omega)$, which describes the response of $s_{\downarrow\uparrow}(q)=\sum_{k}c^\dagger_{k\downarrow}c_{k+q\uparrow}$ to a perturbation driven by its conjugate field. Within RPA~\cite{Kubo,Sandri} one obtains
\be
\chi_{\downarrow\uparrow}(q,\omega)=\frac{\chi^0_{\downarrow\uparrow}(q,\omega)}{1+g\chi^0_{\downarrow\uparrow}(q,\omega)},
\ee
with
\be
\chi^0_{\downarrow\uparrow}(q,\omega)=\int\frac{d^3k}{(2\pi)^3}\frac{n_\uparrow(k)-n_\downarrow(k+q)}{\omega+\xi_\uparrow(k)-\xi_\downarrow(k+q)},
\ee
\noindent where, $\xi_{\sigma}(k)=k^2/2m+gn_{\bar{\sigma}}$.  
The splitting between the dispersion of majority and minority spins is $\Delta=g(n_\uparrow-n_\downarrow)$, and, correspondingly, the difference in chemical potentials reads as $\delta\mu\equiv \mu_\uparrow-\mu_\downarrow=\delta \epsilon_F-\Delta$.
\begin{figure}
\includegraphics[scale=0.7]{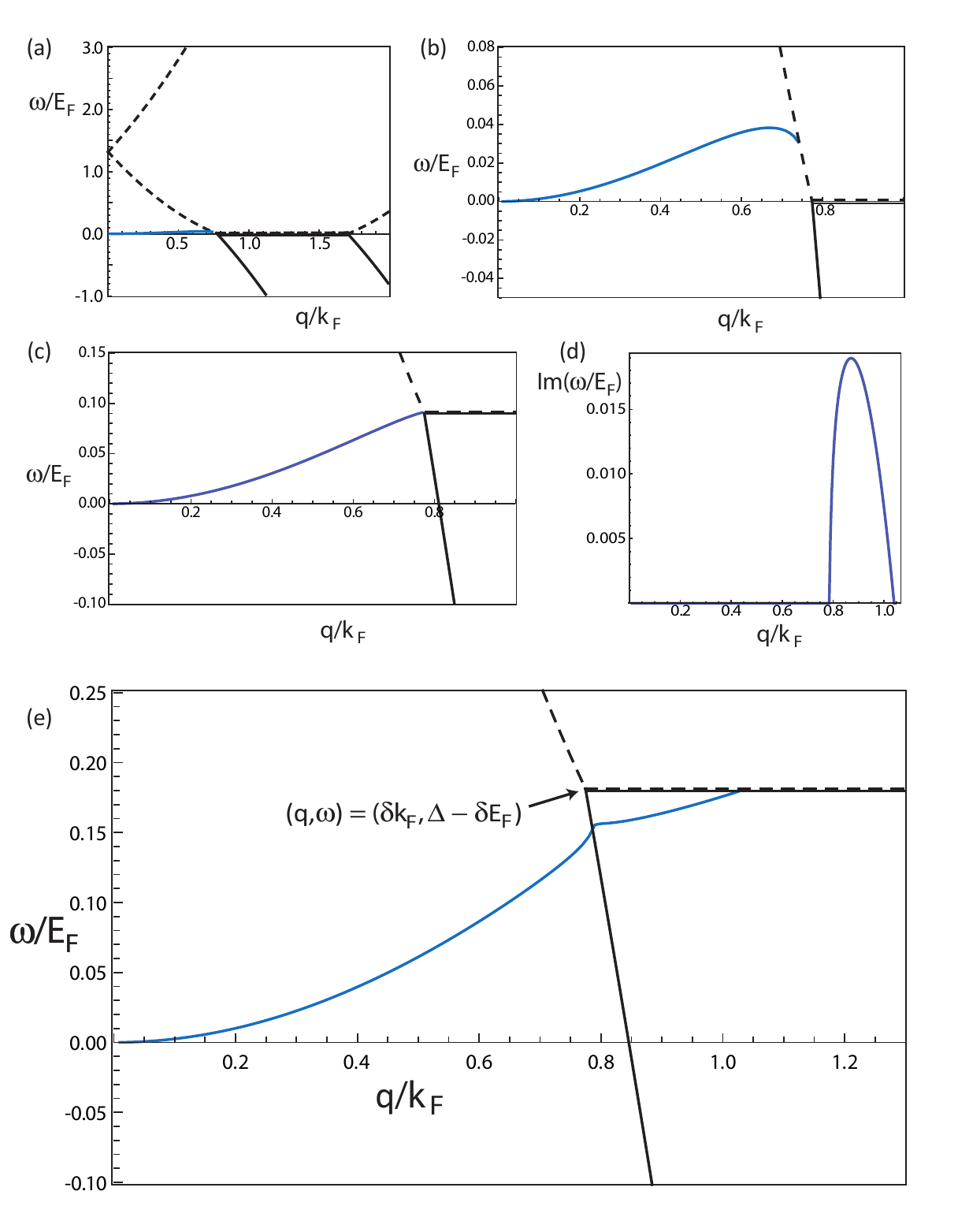}
\caption{\label{spinwaves}(color online) (a) Equilibrium ({\it i.e.} $\delta\mu=0$) spin wave dispersion and particle-hole continuum for $\delta=0.9$. The boundary of the particle-hole excitation energies of majority into minority corresponds to the dashed line, the solid lines bound the negative of the energies of minority into majority excitations, and the blue line is the dispersion of spin waves. (b) Zoom in of Fig.~(a). The spin waves cease to exist when they become degenerate with particle-hole excitations of majority into minority spin, which allows them to decay into particle-hole excitations. (c) Spin wave dispersion at the critical repulsion $g_{crit}$ for the onset of transverse instabilities for $\delta=0.9$. (d) and (e) correspondingly show the imaginary (growth rate of instability) and real parts of the spin wave dispersion at a repulsion strength $g\nu_0=1.25$ for $\delta=0.9$. The instability appears when the spin waves have minus the energy of minority-into-majority particle-hole excitations as explained in the text, and it would look like a spin-density wave of transverse magnetization with a spatial modulation corresponding to the inverse wavevector of the fastest growing mode $\sim q^{-1}_{max}$ during the initial dynamics.}
\end{figure}

Collective modes are found by solving $1+g\chi^0_{\downarrow\uparrow}(q,\omega)=0$. This same equation determines the dispersion of spin waves in a ferromagnet in equilibrium~\cite{Kubo}. In Fig.~\ref{transverse} we depict the critical repulsion strength $g_{\textrm{crit}}$ at which an instability in the transverse modes appears. Since no unstable modes exist in equilibrium, the critical repulsion strength for the appearance of unstable collective modes for a given polarization must be larger than the interaction strength that is needed to sustain such polarization in equilibrium, {\it i.e.} $g_{\textrm{crit}}>g_{\textrm{eq}}(\delta)$.

We find that the unstable modes have a finite momentum $q \ge k_{F\uparrow}-k_{F\downarrow}$, and that they evolve continuously from the conventional spin-waves that the system would have in equilibrium. The instability exists whenever the spin wave spectrum, $E_{\textrm{sw}}(q)$, coincides with minus the energy of a particle-hole excitation of minority into majority spin, {\it i.e.}  when $E_{\textrm{sw}}(q)=-(\xi_\uparrow(k+q)-\xi_\downarrow(k))$, for some $k$ such that $k<k_{F\downarrow},k_{F\uparrow}<|k+q|$. This can be seen as a necessary condition for the spontaneous excitation of magnons and particle-hole pairs while conserving total energy, momentum and spin. It is thus analogous to the critical interaction for pair formation discussed in the BEC side of the resonance, but instead of the particle-particle binding to form a molecule we consider particle-hole binding to form a magnon of spin $-1$ which leaves a particle-hole excitation in the polarized Fermi sea of spin $+1$. The absence of unstable modes at long-wavelengths is thus a consequence of the Pauli exclusion principle which prohibits minority-into-majority excitations of momenta $q<k_{F\uparrow}-k_{F\downarrow}$~\cite{Conduit}. Fig.~\ref{spinwaves} illustrates how the spin wave spectrum evolves from being damped at equilibrium to unstable beyond the critical interaction strength.

\section{Summary}\label{summary}

Our results imply that one should expect considerable differences in the initial unstable dynamics of polarized and unpolarized gases of ultracold fermions, and that features of the initial unstable state can be significantly different from those expected from the equilibrium phase diagram in the population imbalance-scattering length coordinates.

When the resonance is approached from the BCS side, finite momentum pairing instabilities and density-longitudinal spin instabilities are encountered. Pairing at finite momentum is the fastest growing pairing instability over a wide range of parameters. This should be contrasted with the equilibrium FFLO phase region, conjectured to occupy a rather small portion of the phase diagram~\cite{theoimbalance}. The density-longitudinal spin instability can be viewed as a precursor of the phase separation between a balanced superfluid and excess majority fermions, expected to settle in at longer time scales. We predict an evolution of unstable modes as the population imbalance increases, with a transition between finite momentum pairing at small polarizations ($\delta \lesssim 0.4$), and phase-separation modes at larger polarizations.

When the resonance is approached from the BEC side, we find that the zero-momentum pairing instability always occurs before the longitudinal spin-density instability. The latter instability evolves from what would be the Stoner instability in the unpolarized gas. This finding seems discouraging for simulating a Stoner-like transition in a polarized gas of fermions. Nevertheless, as illustrated with a simplified model in Sec.~\ref{sectransverse}, instabilities of the transverse spin modes are also expected to appear before the longitudinal spin and density instabilities. Therefore, an analysis of these instabilities which systematically accounts for the modification of the interactions near the resonance is necessary to resolve the fate of the competition between the transverse-spin and pairing instabilities at finite polarizations.

It is worth emphasizing that in a polarized gas with globally conserved populations of each spin species, some of the physics of itinerant electron ferromagnetism can be explored, independently of the presence of spin-density instabilities. For example, in our simple model of Sec.~\ref{sectransverse}, the spectrum of spin waves acquires a positive mass (magnetization stiffness) for interaction strengths below any spin instability, and below the interaction strength that would spontaneously sustain the population imbalance in equilibrium. Experimental studies of spin-wave dynamics of fully polarized states close to a Feshbach resonance are likely to shed light on many-electron physics questions that arise in the context of itinerant electron magnetism, and also on pairing physics questions that are unique to the cold atom problem.

\begin{acknowledgments}
The authors wish to thank Mehrtash Babadi, Eugene Demler, Rembert Duine, David Pekker, Rajdeep Sensarma and Joseph Thywissen for valuable discussions. This work was supported by the Welch Foundation under Grant No. TBF1473.
\end{acknowledgments}

\end{document}